**2D antiscatter grid and scatter sampling based CBCT pipeline for image guided radiation therapy**

**Authors:** Farhang Bayat[a], Dan Ruan[b], Moyed Miften[a], Cem Altunbas[a]

[a] Department of Radiation Oncology, University of Colorado School of Medicine, Aurora, CO 80045, USA.

[b] Department of Radiation Oncology, University of California Los Angeles, Los Angeles, CA 90095, USA.

**Corresponding Authors:**

1. Farhang Bayat

   Email address: farhang.bayat@cuanschutz.edu

   Postal Address: 1665 Aurora Court, Suite 1032 | MS F-706 | Aurora, CO 80045

2. Cem Altunbas

   Email address: cem.altunbas@cuanschutz.edu

   Postal Address: 1665 Aurora Court, Suite 1032 | MS F-706 | Aurora, CO 80045

## Abstract

**Background**: Poor tissue visualization and quantitative accuracy in cone beam CT (CBCT) is a major barrier in expanding the role of CBCT imaging from target localization to quantitative treatment monitoring and plan adaptations in radiation therapy sessions. While CBCT image quality problem mostly emanates from high scatter fluence, scatter suppression by itself is not sufficient to achieve high CBCT image quality.

**Purpose:** To further improve image quality in CBCT, 2D antiscatter grid-based scatter rejection was combined with a raw data processing pipeline and iterative image reconstruction. The culmination of these steps was referred as quantitative CBCT (qCBCT), and its performance was evaluated in a comprehensive set of experiments.

**Methods:** qCBCT data processing steps include 2D antiscatter grid implementation, measurement-based residual scatter, image lag, and beam hardening correction for offset detector geometry CBCT with a bow tie filter. Images were reconstructed with iterative image reconstruction to reduce image noise. To evaluate image quality, qCBCT acquisitions were performed using a variety of phantoms to investigate the effect of object size and its composition on image quality. qCBCT image quality was benchmarked against clinical CBCT and multi-detector CT (MDCT) images.

**Results:** Addition of image lag and beam hardening correction to scatter suppression reduced HU degradation in qCBCT by 10 HU and 40 HU, respectively. When compared to gold standard MDCT, mean HU errors in qCBCT and clinical CBCT were 10±12 HU and 27±27 HU, respectively. HU inaccuracy due to change in phantom size was 22 HU and 85 HU in qCBCT and clinical CBCT images, respectively. With iterative reconstruction, contrast-to-noise ratio improved by a factor of 1.25 when compared to clinical CBCT protocols.

**Conclusions**: Robust artifact and noise suppression in qCBCT images can reduce the image quality gap between CBCT and MDCT, improving qCBCT's promise in fulfilling the tasks that demand high- quantitative accuracy, such as CBCT-based dose calculations and treatment response assessment in image guided radiation therapy.

## Keywords

## 1. Introduction

Even though CBCT imaging is the most used 3D in-room imaging modality to localize targets during image guided radiation therapy (IGRT), its poor quantitative accuracy and tissue visualization remains a challenge in implementation of contemporary treatment paradigms, such as CBCT-based dose delivery monitoring and radiation treatment plan adaptations [1]. Moreover, CBCT images acquired during radiation treatments can help to predict radiation therapy outcomes and be potentially used for personalized tailoring of radiation treatments. However, such approaches require accurate CT numbers and clear visualization of image features, which remain as challenges due to relatively poor image quality of CBCT images [2-4].

Reasons behind poor CBCT image quality are numerous, such as scattered radiation, motion artifacts due to long scan times, limitations of flat panel detectors (poor quantum efficiency, image lag, limited dynamic range), suboptimal image reconstruction due to circular source trajectory, and suboptimal beam hardening correction[5]. Cumulative effect of these problems yields poor CT number accuracy, increased noise, lower contrast, and blurring that degrade visualization of anatomical structures and targets.

Among these, scattered radiation is considered one of the most fundamental problems in CBCT, for which we previously proposed and investigated a novel 2D antiscatter grid-based approach to address the issue [6-8]. While this approach can robustly suppress scatter [6,7,9], scatter suppression by itself is not sufficient to improve CBCT image quality to levels comparable to MDCT. Other data correction methods, such as image lag and beam hardening corrections, are needed. Furthermore, scatter correction and relatively lower quantum efficiency and higher electronic noise of amorphous silicon flat panel detectors result in increased image noise in CBCT images [6,10,11]. If data correction and image denoising methods were to be combined with 2D antiscatter grid-based scatter suppression in one data processing pipeline, both HU accuracy and soft tissue visualization in CBCT imaging can be potentially improved further.

In this work, we developed a raw data processing and iterative reconstruction pipeline that is used in conjunction with a 2D antiscatter grid for high fidelity CBCT imaging in radiation therapy. The pipeline consists of a 2D antiscatter grid prototype [8,12], a measurement-based scatter correction method [6,13], image lag correction, and beam hardening correction. In addition, an iterative image reconstruction method was incorporated as part of the pipeline to reduce image noise. This approach is referred to as quantitative CBCT, or qCBCT.

Image quality improvement methods in CBCT have been extensively studied [5], and state-of-the-art clinical CBCT systems combine numerous image quality improvement methods in one data processing pipeline [14-17]. However, none of the existing work utilizes robust hardware-based scatter suppression in conjunction with software-based data correction and image denoising methods.

In short, the novelty of this work lies in the integration of robust hardware-based scatter suppression, raw data correction, and iterative image reconstruction in one data processing pipeline for CBCT guided radiation therapy. To quantify the image quality improvement, benchmarking comparisons were made with respect to the current state of the art in clinical CBCT images used in IGRT and MDCT images utilized in radiation therapy treatment planning.

## 2. Methods

### 2.1. Overview of data processing steps and image reconstruction

Fig.1 presents the overall flow of our data processing pipeline with details of each step explained in the following sections. Briefly, a CBCT scan is acquired with a 2D antiscatter grid in place, and by using a linac mounted CBCT system. CBCT projections are dark and flat field-corrected by the clinical CBCT system and subsequently exported for downstream processing, which are referred as raw projections.

Each raw projection was corrected for image lag, residual scatter, grid septal shadows (referred as gain correction), and beam hardening. For objects larger than transverse field of view, a truncation correction was also applied [18]. CBCT images were reconstructed either by using filtered backprojection or applying OS-ASD-POCS [19] iterative reconstruction method, both in TIGRE image reconstruction toolkit [20]. For filtered backprojection, an offset detector weighting scheme was implemented to reconstruct in offset detector geometry [21].

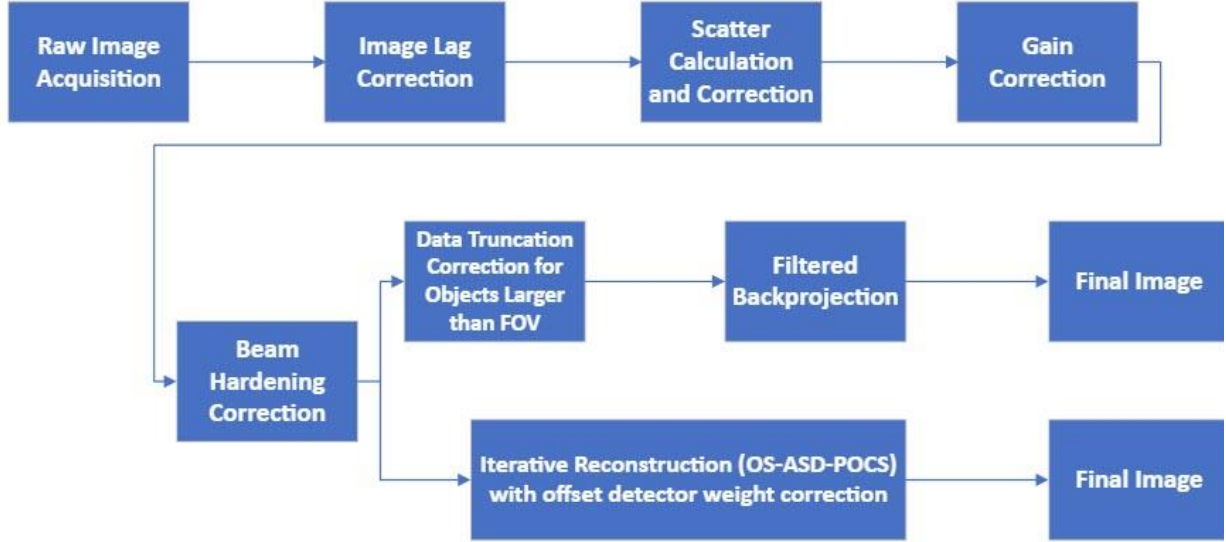

*Fig.1. Raw Data Correction and Iterative Reconstruction Pipeline for qCBCT scans acquired with 2D antiscatter grid.*

## 2.2. Gain Correction

Gain correction is a flat field correction procedure that suppresses the 2D grid's septal shadows in projections and reduces grid-induced artifacts in CBCT images. Gain correction starts with the acquisition of a flood CBCT scan, i.e., a scan without a phantom, to characterize grid shadows in projections as a function of source or gantry angle [22]. From these flood projections, gain maps are generated as below:

$$GM(x, y, \varphi) = \frac{I_0}{Flood\ projections(x, y, \varphi)} \tag{1}$$

Where $x$ and $y$ represent detector pixel indices in each dimension while $\varphi$ represents projection index (scan angle). Subsequently, gain maps are multiplied with phantom CBCT projections at matching source angles to suppress 2D grid's septal shadows in projections.

## 2.3. Image Lag Correction

Each CBCT projection in a scan inherits a residual signal from prior CBCT projections in the scan, a phenomenon known as image lag [23]. If the image lag is unaccounted for, it may cause image artifacts and degrade HU accuracy. For image lag correction, an empirical method was implemented [24]. A flood CBCT scan was acquired, and a second order exponential model was fitted to parameterize the mean signal intensity as a function of projection number in the scan, $I_{flood}^{fitted}(i), i = 1 \ldots, M$. Relative intensity difference between two consecutive flood projections yielded the relative image lag, referred as lag coefficient,

$$coeff_{flood}^{lag}(i) = \frac{I_{flood}^{fitted}(i+1) - I_{flood}^{fitted}(i)}{I_{flood}^{fitted}(i+1)}, i = 1 \ldots, M-1 \tag{2}$$

Subsequently, flood and phantom CBCT projections were corrected using the following,

$$Flood_{corrected}(i)$$
$$= Flood_{uncorrected}(i) - Flood_{uncorrected}(i-1) \times coeff_{flood}^{lag}(1)$$
$$- Flood_{uncorrected}(i-2) \times coeff_{flood}^{lag}(2) - \cdots$$
$$- Flood_{uncorrected}(i-k) \times coeff_{flood}^{lag}(k) \tag{3}$$

$$Phan_{corrected}(i)$$
$$= Phan_{uncorrected}(i) - Phan_{uncorrected}(i-1) \times coeff_{flood}^{lag}(1)$$
$$- Phan_{uncorrected}(i-2) \times coeff_{flood}^{lag}(2) - \cdots$$
$$- Phan_{uncorrected}(i-k) \times coeff_{flood}^{lag}(k), k = \min(i-1, M) \tag{4}$$

where $Phan_{uncorrected}(i-k) \times coeff_{flood}^{lag}(k)$ represents the lag carried over from the frame $i-k$ to $i$. Since magnitude of image lag from one projection to subsequent projections goes down exponentially, only image lag up to M prior projections were accounted using this correction strategy, thus the choice between minimum of $i-1$ and M in above equation. Flood CBCT scan to characterize image lag was acquired without a 2D antiscatter grid. In our application, M = 300 is used.

## 2.4. Grid-Based Scatter Sampling (GSS) Method

While 2D grid rejects majority of scatter fluence, it does not fully eliminate scatter. Therefore, residual scatter in projections must be corrected. Previously developed GSS method was employed to correct residual scatter as described in prior publications [6,13]. For completeness, a brief description of the method is provided below. In the GSS method, a 2D antiscatter grid is employed as a residual scatter measurement device. The grid's footprint, or septal shadow, acts as micro- fluence modulators, where ratio of signal in grid holes to shadows in a small neighborhood of pixels varies as a function of scatter intensity. In gain corrected projections, this variation in signal ratio manifests itself as a signal intensity difference, $d$, between pixels residing in grid shadows and adjacent pixels residing in grid holes. Assuming scatter intensity $S$ is piecewise uniform in pixels residing both in grid shadows and grid holes in a small neighborhood of pixels (typically a 7×7 pixel region, corresponding to an area of 2.7×2.7 mm$^2$), I can be defined as [6,13],

$$S(x_1, y_1) = \frac{d(x_1, y_1)}{GM_{grid}(x_1, y_1) - GM_{hole}(x_2, y_2)}. \tag{5}$$

Where $x_1$ and $y_1$ are for pixels in grid shadows and $x_2$ and $y_2$ are for pixels in grid holes. $GM_{grid}$ and $GM_{hole}$ are the values of gain maps in grid septal shadows and holes, respectively.

Using Eq. 5, residual scatter was first estimated in pixels residing in grid shadows and subsequently, residual scatter values in each detector pixel were obtained via interpolation. Residual scatter was subtracted from projections to achieve scatter corrected projections.

## 2.5. Beam Hardening Correction

For polyenergetic x-rays, the log attenuation projections can be calculated by

$$p = -\log(\frac{I}{I_0}) = -\log(\int \Omega(E) e^{-\int \mu_{E,s} ds} dE) \tag{6}$$

Where $\Omega(E)$ is the incident x-ray spectrum, $I_0$ and $I$ represent the incident and transmitted intensities and $\mu_{E,s}$ is the linear attenuation coefficient of the object at a specific energy $E$ and path $s$.

The relationship between $p$ and the path $s$ is not linear, a problem known as beam hardening. In the clinical CBCT system, beam hardening is caused by both the imaged object and the aluminum bow tie filter [25]. Specifically, effective thickness of the bow tie filter varies between 3-27 mm in the transverse direction, which generates noticeable beam hardening artifacts in the transverse direction in CBCT images (Fig. 2). To address this issue, a water-equivalent beam hardening correction method was implemented that also reduced bow tie filter induced beam hardening artifacts [25]. The goal of this method is to generate a mapping function, such that the projection log attenuation values have a linear relationship with water equivalent thickness of the object (path length). To calculate effects of beam hardening, source spectra at 125 kVp and 140 kVp with 0.89 mm Titanium filtration were used [26,27]. For any water equivalent path length $s$ with bowtie thickness $l$ a correction factor $CF(s,l)$ was calculated and log attenuation was updated:

$$p^*(s,l) = p(s,l) \times CF(s,l) \tag{7}$$

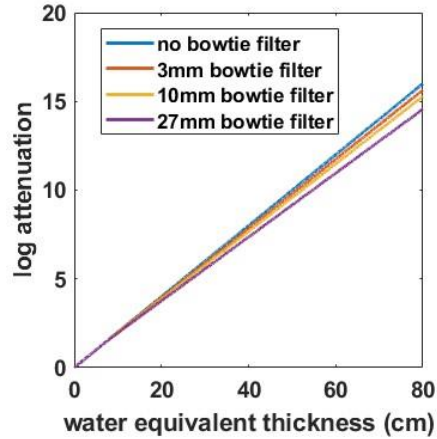

*Fig.2. Effect of bow tie filter thickness on projection log attenuation values.*

## 2.6. Iterative Reconstruction

An iterative reconstruction approach was adopted that employed Ordered Subset Adaptive Steepest Descent Projection Onto Convex Sets (OS-ASD-POCS) algorithm[19] to seek the optimal image $\vec{f}$ with respect to the following regularized L1 norm objective [28]:

$$\vec{f}* = argmin \left| A\,\vec{f} - \tilde{g}_{data} \right| + \tau ||\vec{f}||_{TV} \tag{8}$$

The first data fidelity term encourages consistency between measured projections and reconstructed image, where $A$ is the system matrix and $\tilde{g}_{data}$ is the acquired post-correction CBCT projection. The second term penalizes total variation to encourage piecewise smoothness in the reconstructed image. The weight constant $\tau$ controls the tradeoff between these two energies. The implementation is performed with the TIGRE MATLAB/CUDA toolbox with GPU acceleration [20].

### 2.6.1. Implementation of OS- ASD-POCS in offset detector geometry

We observed that OS-ASD-POCS implementation may cause ring and streak artifacts in offset detector geometry (Fig. 3). These artifacts are caused by data inconsistency induced by scatter on the medial edge of the detector, where projection data is always truncated due to 16 cm lateral offset in detector position (Fig. 4a). To demonstrate the data inconsistency problem, CBCT scans were acquired with 1D antiscatter grid, 2D antiscatter grid, and with 2D antiscatter grid with GSS scatter correction. They were reconstructed using the OS-ASD-POCS method (Fig. 3). With 1D grid, residual scatter in projections was relatively high, and associated data inconsistency artifacts were severe. When 2D antiscatter grid was used, scatter fluence and data inconsistency artifacts were reduced substantially. With additional scatter correction, artifacts were reduced further, but not fully eliminated. Truncated edge of the projection back projects to the location of ring artifact, which is 4 cm away from the piercing point.

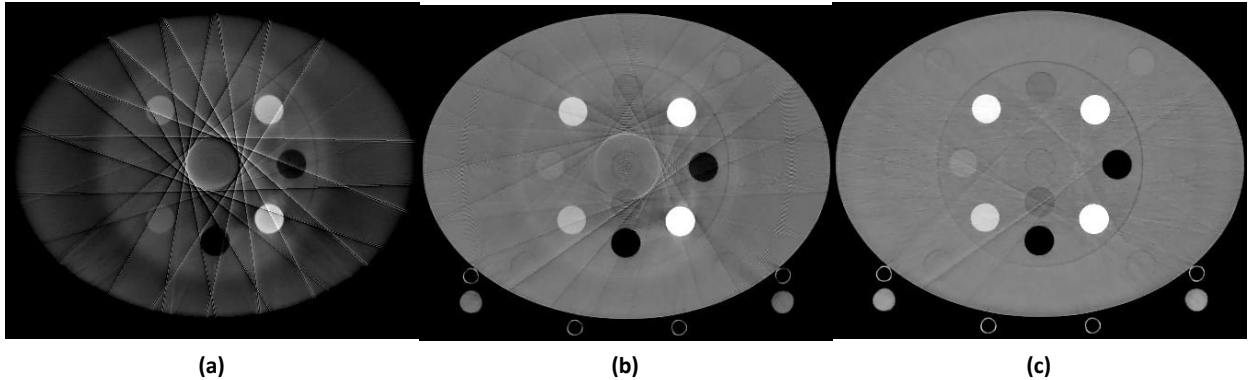

(a)                                             (b)                                             (c)

*Fig.3. CBCT scans were acquired using offset detector geometry and reconstructed using OS-ASD-POCS with (a) 1D antiscatter grid (b) 2D antiscatter grid (c) 2D antiscatter grid + GSS scatter correction. When scatter is present, projection truncation due to offset detector geometry causes ring and streak artifacts due to data inconsistency at the truncated edge of the detector. The use of 2D grid instead of 1D grid reduces these artifacts. 2D grid and residual scatter correction substantially reduce artifacts, but artifacts are not fully eliminated. HU window= [-500 500].*

To address this issue, a projection-domain weighting scheme was implemented, where pixel specific values of the data fidelity term were gradually reduced by applying multiplicative weights to pixels-by-pixel difference between experimentally acquired projection and forward projected image reconstruction (Fig. 4b). These projection weights were identical to offset detector weights used in filtered backprojection reconstruction [21]. These weights for an $R$ source to origin-distance, $t$ spatial position of the detector and $\theta$ the corresponding smaller span of the array (all presented in Fig. 4a), are as below:

$$w(t) = \frac{1}{2}\left( \sin\left( \frac{\pi \tan^{-1} \frac{t}{R}}{2 \tan^{-1} \frac{\theta}{R}} \right) + 1 \right), \qquad -\theta \leq t \leq \theta \tag{9}$$

It is important to emphasize that the purpose of offset detector weights in iterative reconstruction is different than the ones in filtered backprojection, where offset detector weights reduce the weight of detector pixels that are double sampled due to offset detector geometry.

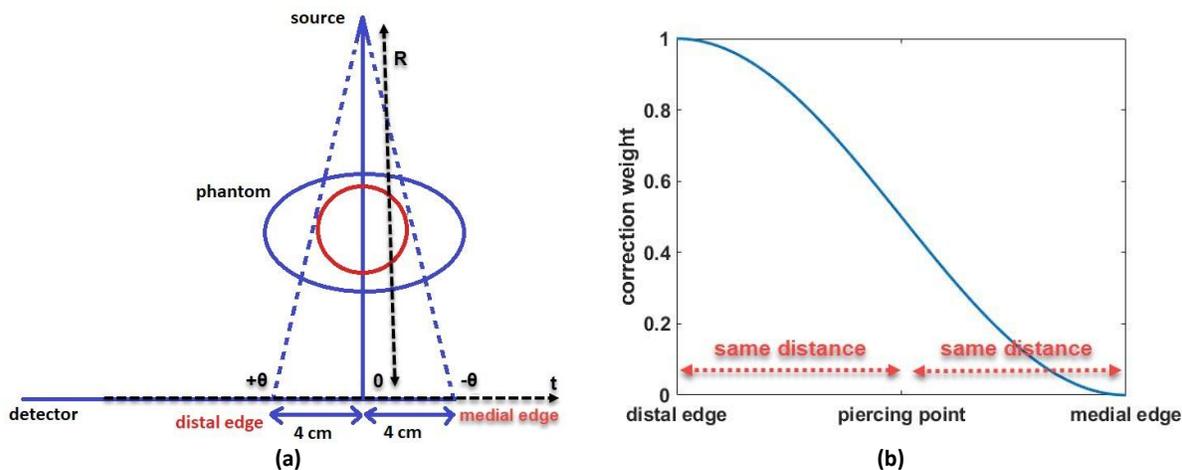

*Fig.4. (a), Offset detector geometry used in the CBCT experiments. Medial edge of the projection is closest to the piercing point and the object is truncated, which causes data inconsistency when scatter is present. (b), data inconsistency is reduced by applying correction weights to the fidelity term of pixels within 4 cm of the piercing point.*

In iterative reconstruction, offset detector weights were applied to modulate the data fidelity term, by multiplying the projection residual, which was subsequently back projected to the image domain to be denoised via TV minimization to complete one cycle of the iterative reconstruction (Fig. 5).

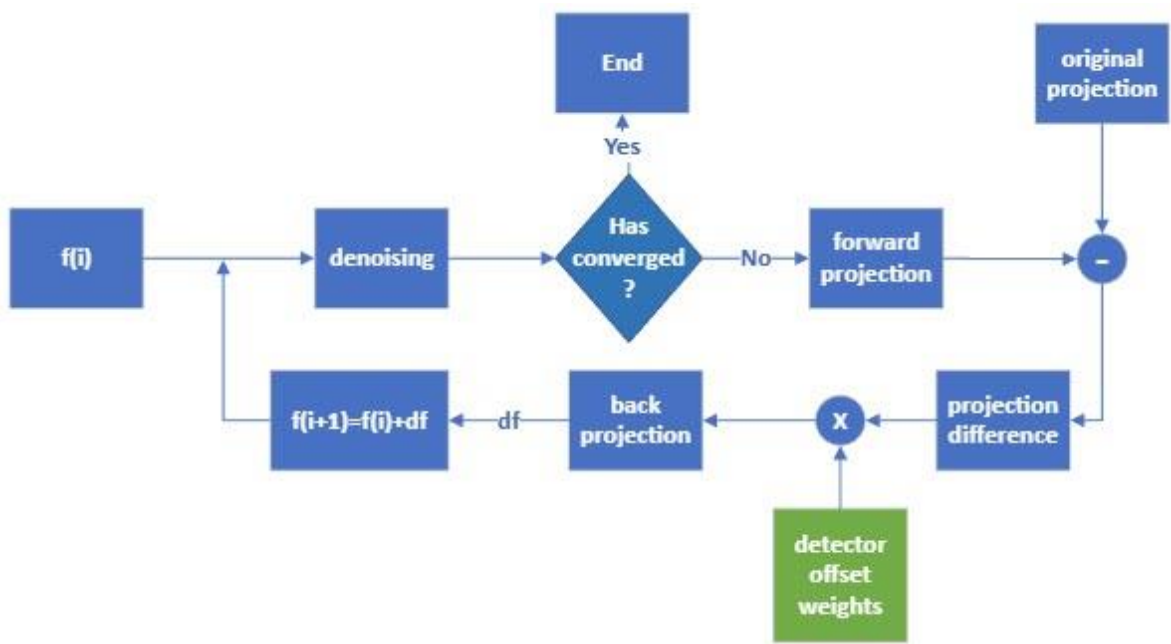

*Fig.5. Flowchart for updated iterative reconstruction step to account for scatter induced data inconsistency and associated ring artifacts.*

## 2.7. Image quality evaluation metrics

Several HU accuracy evaluation methods and CNR were used in image quality evaluations of qCBCT and benchmarked against clinical CBCT and MDCT images. Since image quality can vary substantially depending on the location of regions of interest (ROI) within a phantom image, all CBCT and MDCT image sets of any given phantom were rigidly co-registered before image quality evaluations.

### 2.7.1. HU Loss

HU loss represents how HU values degrade in images when the size of the object changes. HU loss was evaluated using 2 phantoms made from the same material but in two different sizes,

$$\Delta HU_{method} = \left| HU_{small,method} - HU_{large,method} \right| \tag{10}$$

$\Delta HU_{method}$ is the average absolute HU difference between any method's HU values for the large and the small object amide from the same material.

HU loss depends not only on the object size, but also on the material properties. ROI locations chosen for analyses were separated into two groups, one for soft tissue like regions and the other for bone-like regions in each phantom.

### 2.7.2. Relative contrast-to-noise ratio improvement factor (rCNR)

Relative change in CNR (rCNR) with respect to Clinical FDK images of the same phantom was calculated as,

$$rCNR_{phantom,ROI(i)}^{method} = \frac{CNR_{phantom,ROI(i)}^{method}}{CNR_{phantom,ROI(i)}^{TrueBeam\ FDK}} \tag{11}$$

Where $CNR_{phantom,ROI(i)}^{method}$ represents the calculated CNR value for the data processing method of interest in $ROI(i)$ for each phantom. The ROIs were selected to span contrast objects in electron density and Catphan phantoms.

### 2.7.3. HU nonuniformity

To calculate average HU nonuniformity, a total of 20 ROIs were evenly distributed and selected in each transverse CBCT image's soft tissue-equivalent background. Mean HU nonuniformity for each ROI was calculated as:

$$nonuniformity = \frac{1}{N}\sum_{i=1}^{N} |\overline{HU}\big(ROI(i)\big) - \{\frac{1}{N}\sum_{i=1}^{N} \overline{HU}\big(ROI(i)\big)\}|$$

Where $\overline{HU}\big(ROI(i)\big)$ is the average HU calculated in $i^{th}$ ROI placed in the same phantom material. Mean HU nonuniformity was defined as the absolute difference between the HU value of each ROI with respect to the ensemble mean of all ROIs in each phantom image. Ideally, nonuniformity is expected to be 0 HU.

**2.7.4. Modulation Transfer Function (MTF) and bar pattern analysis**

Spatial resolution in terms of MTF and line-pair response was further quantified to complement CNR as image quality metrics. To measure MTF, one of the high contrast objects in the head-sized electron density phantom was used, and MTF was calculated using the method by Friedman et al[29]. MTF level at 10% was used to compare spatial resolution. In addition, line-pair pattern module in small Catphan phantom was evaluated.

**2.7.5. HU correlation histogram**

To fully depict the HU deviations across the whole volume of a given image quality phantom, beyond zeroth and first order statistics with HU loss and nonuniformity, the correlation of HU values between CBCT and MDCT images was investigated across the whole imaged volume of the phantom. While a high level of correlation between CBCT and MDCT HU values is desired, establishing a reference HU correlation between CBCT and MDCT HU is challenging due to HU inaccuracies in CBCT images. In this work, head-sized electron density phantom images were utilized to estimate the *reference* correlation which were affected the least from scatter, beam hardening and image lag problems. In co-registered CBCT and MDCT images, an HU correlation histogram was calculated by pairing CBCT and MDCT image voxels. As a first order approximation, the correlation between the HU values of CBCT and MDCT was assumed to be piecewise linear. Two separate linear fits were performed for CBCT-MDCT HU pairs, one for soft tissue-like materials and the other for bone-like materials. Root mean square error (RMSE) between the HU values of CBCT images and the linear fit was calculated to quantify the CBCT HU inaccuracy and referred to as histogram RMSE (h-RMSE). Subsequently, the linear fit extracted from head-sized phantom images was used for calculating h-RMSE-based HU inaccuracy in CBCT images of other phantoms.

**2.8. Experiment Setup**

Two types of CBCT data were acquired using a TrueBeam linac mounted CBCT (Varian Medical Systems, Palo Alto, CA). First data set was acquired with our 2D antiscatter grid prototype and the proposed data processing pipeline, referred as "qCBCT". A second data set was acquired using a standard "Pelvis" CBCT protocol in the system, referred as "Clinical CBCT".

Clinical CBCT scans were reconstructed using two different options [30].1) Standard reconstruction that employs filtered backprojection similar to Feldkamp-Davis-Kress algorithm [31] and scatter correction based on scatter kernel superposition [15]. This method was referred as Clinical FDK. 2) A statistical reconstruction method combined with a more robust scatter correction algorithm that solves Boltzmann transport equation [14]. This method was referred as Clinical iCBCT.

Likewise, qCBCT images were reconstructed using FDK algorithm and iterative reconstruction, which were referred as qCBCT FDK and qCBCT IR, respectively**.** In our image reconstruction using OS-ASD-POCS, zero-value image was used for initialization. In each iteration, a subset of $blocksize$ =50 projection views were used to approximate the gradient of the cost function. We count one iteration after all 900 projections were all used with $900/50 = 18$ sub-iterations. Overall, $iter = 50$ POCS/descent iterations were used. In each iteration, the image update coefficient factor set to be $\lambda = 1$ without any reduction factor for a faster convergence. The TV-steepest descent step size was initialized to $\alpha = 0.002$ and executed for $TV_{iter} = 15$ times when the change in the image due to TV-steepest descent is assessed. If this change exceeded $r_{max} = 0.94$ times data fidelity POCS based update, the TV-steepest descent step size is reduced by $\alpha_{red} = 0.95$.

Both qCBCT and clinical datasets were acquired using the same Pelvis CBCT protocol parameters in offset detector geometry (i.e., half fan geometry). 900 projections were acquired with bow tie filter and 0.89 mm titanium beam filter in place, and detector pixel size was 0.388×0.388 mm$^2$. Most of CBCT scans were acquired at 125 kVp and 1080 mAs, except pelvis phantoms which were acquired at 140kVp and 1680 mAs. qCBCT images were reconstructed using 0.9×0.9×2 mm$^3$ voxel size, whereas clinical images have a voxel size 0.908×0.908×1.98 mm$^3$.

Focused tungsten 2D antiscatter grid prototype had a grid ratio of 12, grid pitch of 2 mm, and septal thickness of 0.1 mm, designed for offset detector CBCT geometry in TrueBeam [32]. The 2D grid was installed on the FPD, after removing the default (1D) antiscatter grid.

Effects of scattered radiation and CBCT image quality strongly depend on the object size and composition. Therefore, experiments were conducted using 7 different phantoms spanning from 20 cm diameter head sized phantoms to a large pelvis phantom with a lateral dimension of 55 cm. Anatomically more realistic thorax and pelvis phantoms were also employed.

Finally, a third set of image data was acquired with the MDCT (Philips Brilliance Big Bore 16 slice MDCT, Netherlands) which served as the gold standard reference[33]. Images were acquired using 0.9×0.9×3 mm$^3$ voxel size. Slice thickness in MDCT images were larger than CBCT slice thickness to reduce image noise. MDCT scans were acquired in helical mode using 120 kVp and 140 kVp (for pelvis phantom CBCT scans). CTDI values for MDCT images were 16 mGy and 37 mGy at 120 and 140 kVp scans, respectively. Similarly, clinical and qCBCT scans had the same CTDI values, 16 mGy and 37 mGy at 125 and 140 kVp scans, respectively.

## 3. Results

### 3.1. Effect of data processing pipeline on qCBCT image quality

Fig. 6 shows the images of the pelvis electron density phantom obtained at each step of the data processing pipeline. Images acquired with 2D grid only exhibit several different image artifacts. There are prominent shading artifacts across various regions of the phantom. Periphery of the phantom has lower HU values than the center with a clear boundary (blue arrows). This is due to residual scatter and reduced primary in the peripheral regions with bow tie filter, and the subsequent increase in scatter-to-primary ratio (SPR). Residual scatter also caused more prominent dark streaks between high density objects and more severe ring artifacts in the central region (yellow arrow).

While most shading and ring artifacts were suppressed after scatter correction, other artifacts are still visible, such as image lag artifacts in the periphery (black arrow). Central section of the phantom body has higher HU values mostly due to beam hardening introduced by the bow tie filter. After image lag correction, median image HU nonuniformity in the radar artifact region was reduced from 16 to 4 HU (Fig.7a). Likewise, beam hardening correction reduced median HU nonuniformity between the center and periphery from 43 HU to 5 HU (Fig. 7b). In each box plot, the central mark indicates the median, and the bottom and top edges of the box indicate the 25$^{th}$ and 75$^{th}$ percentiles, respectively. The whiskers extend to the most extreme data points not considered outliers, and the outliers (if available) are plotted individually using the '+' marker symbol.

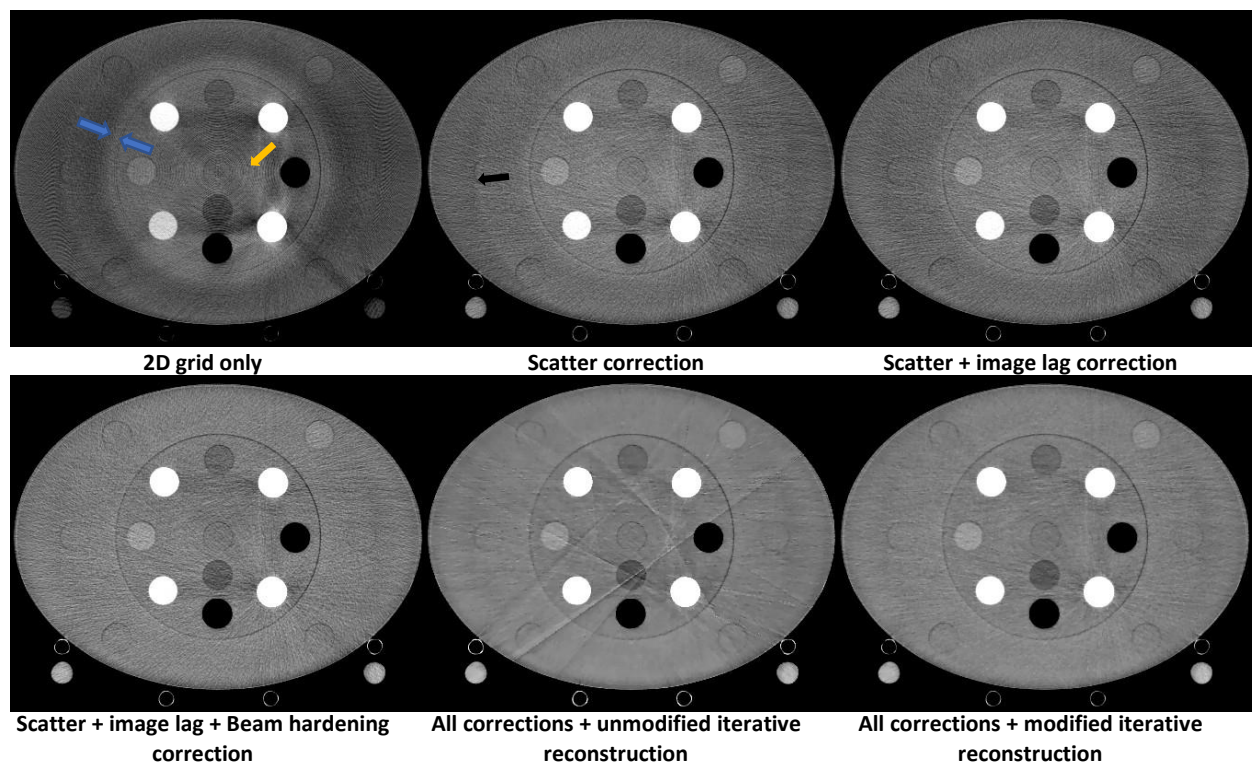

| 2D grid only | Scatter correction | Scatter + image lag correction |

| Scatter + image lag + Beam hardening correction | All corrections + unmodified iterative reconstruction | All corrections + modified iterative reconstruction |

*Fig.6. Step-by-step effect of the data processing pipeline on qCBCT image quality. Blue arrows: transition zone between the peripheral shading artifacts caused by bow tie filter and the central region. Yellow arrow: Ring artifacts caused by residual scatter and associated suboptimal gain correction. Balck arrow: Radar or image lag artifacts. Window = [-250 250]*

Using OS-ASD POCS without offset detector weights reduced stochastic noise, but ring and streak artifacts were present due to data inconsistency at the truncated edge of the detector. After application of offset detector weights, these artifacts were eliminated.

### 3.2. Qualitative evaluation of qCBCT image quality

qCBCT image quality was evaluated in phantoms with different size and compositions (Fig. 8). Image quality differences between qCBCT and clinical images were comparable in head sized electron density and small Catphan phantoms due to lower scatter to primary ratio. Relatively less streak artifacts between high density objects in qCBCT images was also attributed to robust scatter suppression in qCBCT images.

The importance of robust data correction was more evident in larger phantoms. When compared to Clinical FDK, Clinical iCBCT images have less artifacts in large phantoms with heterogenous material composition, such as pelvis electron density and thorax phantoms. qCBCT FDK approach further improves HU accuracy and reduces artifacts across all phantoms investigated. From HU accuracy perspective, qCBCT FDK and IR images have similar appearance, but less noise was observed in qCBCT IR.

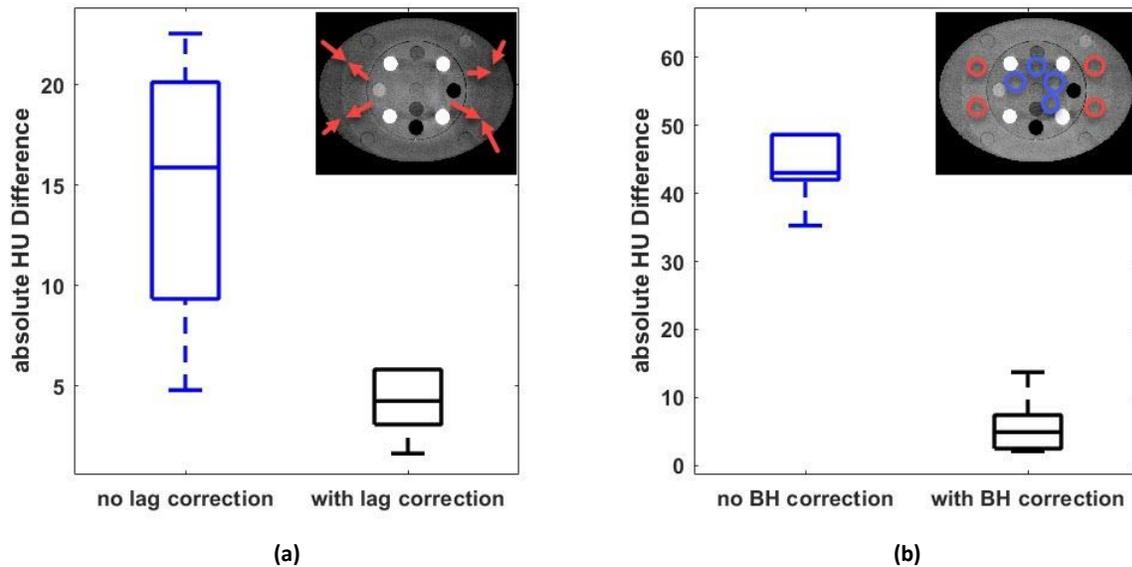

**(a)**     **(b)**

*Fig.7. Effects of image lag and beam hardening correction are show in (a) and (b), respectively. ROI locations for image lag induced HU nonuniformity calculations were indicated by red arrows. ROI locations for image lag induced HU nonuniformity calculations were indicated with circles. Window = [-250 250]*

In visual comparisons of pelvis and large pelvis phantoms (Fig. 8, last two rows), CT number degradation in bony anatomy is particularly evident in Clinical FDK images. While qCBCT significantly reduces such HU degradation, they also appear noisier than Clinical CBCT images. This is due to very low primary fluence transmitted through the large pelvis phantom and relative high residual SPR despite the 2D Grid, which causes substantial noise amplification after scatter correction.

### 3.2.1. HU Loss

Average HU loss in soft tissue equivalent sections of the electron density phantom was 62, 22, 7, and 5 HU for Clinical FDK, iCBCT, qCBCT, and MDCT images, respectively (Fig. 9a). HU loss in qCBCT and MDCT images were similar in electron density phantoms, implying that qCBCT can provide comparable HU accuracy to MDCT. Clinical CBCT images had higher HU loss than qCBCT and MDCT images.

Average HU loss for bone-like objects was significantly higher in all imaging methods; it was 170, 120, 43, and 38 HU for Clinical FDK, iCBCT qCBCT, and MDCT images, respectively (Fig. 9b). Higher HU loss in high density objects was primarily due to higher scatter-to-primary ratio in their projections. Differences in HU loss between qCBCT FDK and IR images were within 1HU.

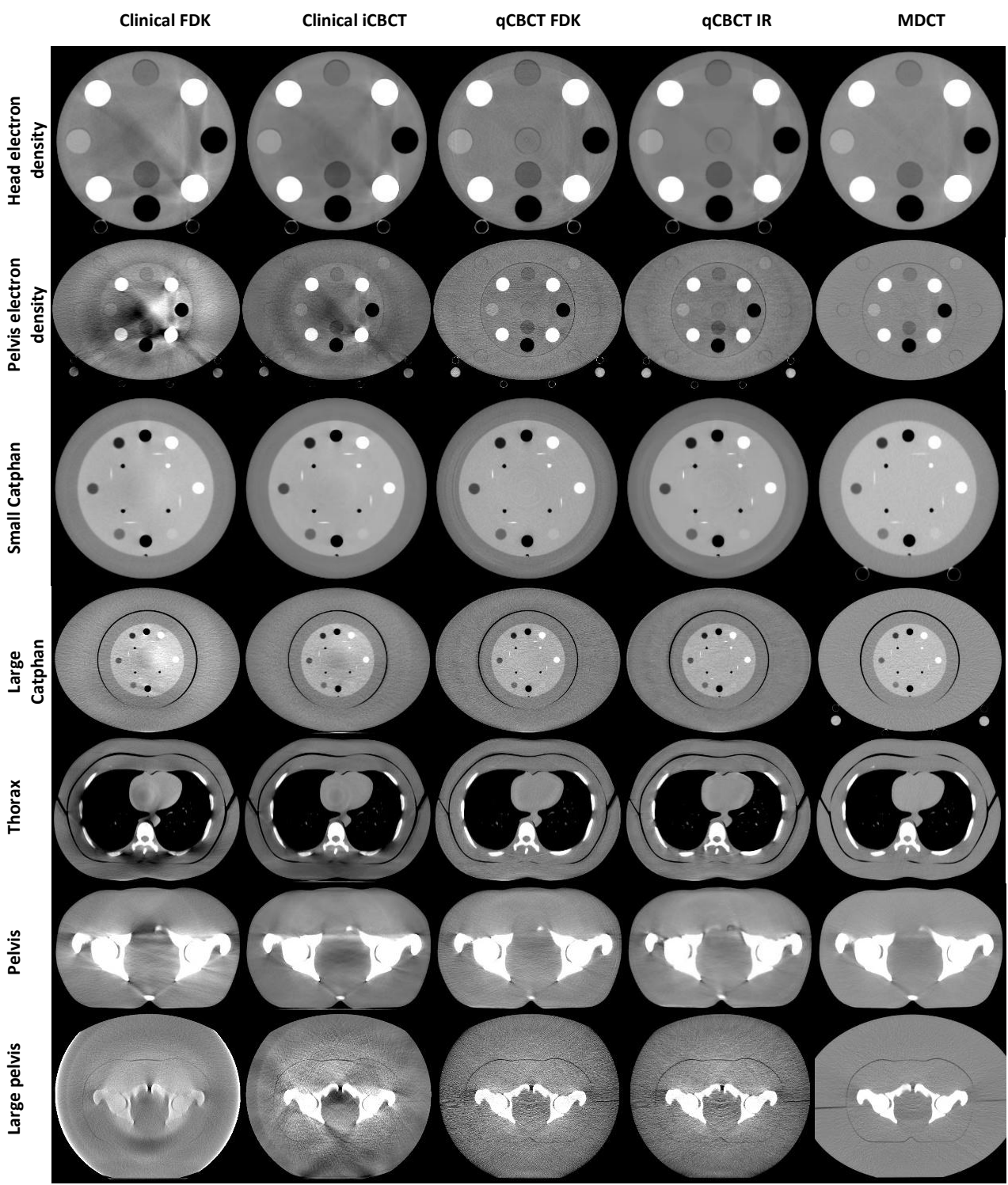

*Fig.8. Image Reconstruction for 7 phantoms using 5 different reconstruction methods. HU windows are [-250 250] for all phantoms except obese pelvis which has a HU window of [-500 500].*

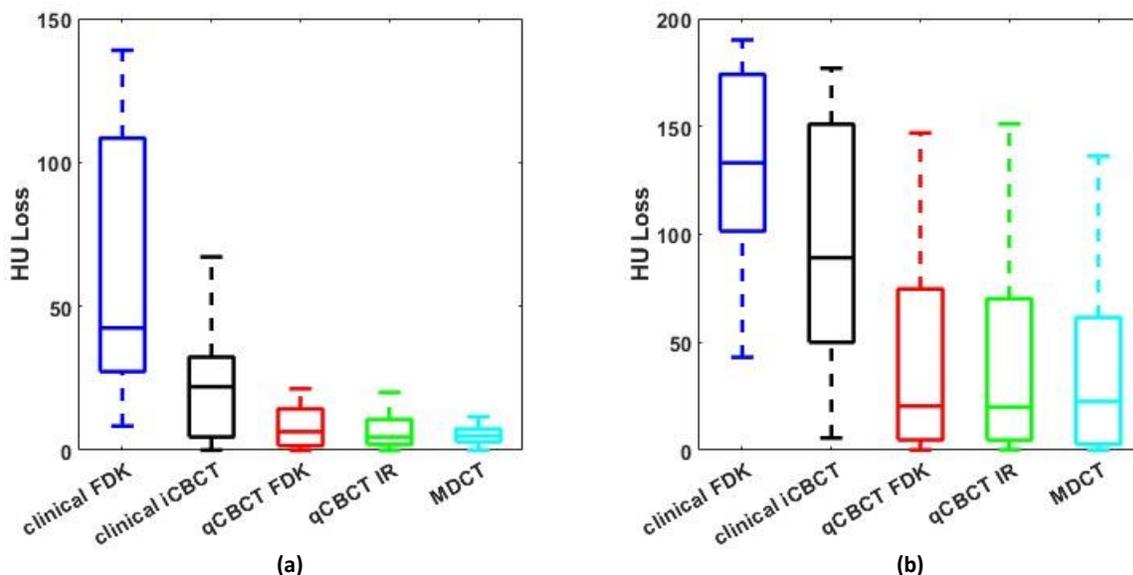

*Fig.9. Boxplots of HU loss as a function of imaging methods, calculated from head and pelvis-sized electron density phantoms in neighborhoods containing (a) water-equivalent background and (b) bone-like objects.*

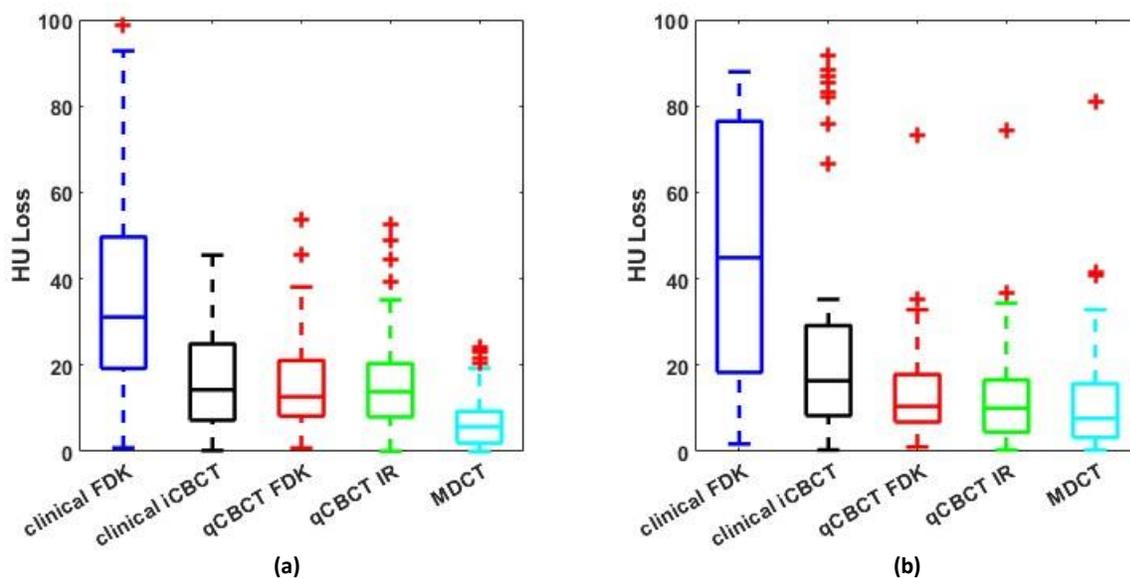

*Fig.10. Boxplots for HU loss using different methods of image reconstruction for small and large Catphan in neighborhoods containing (a) water-equivalent background and (b) contrast objects.*

While similar trends were observed in the Catphan phantom, differences in HU loss were smaller between Clinical and qCBCT images (Fig. 10). Average HU loss in soft tissue equivalent sections was 37, 17, 15, and 7 HU for Clinical FDK, iCBCT, qCBCT, and MDCT images, respectively. Average HU loss in higher contrast objects was 46, 27, 13, and 15 HU for Clinical FDK, iCBCT, qCBCT, and MDCT images, respectively.

### 3.2.2. Relative contrast-to-noise ratio improvement factor (rCNR)

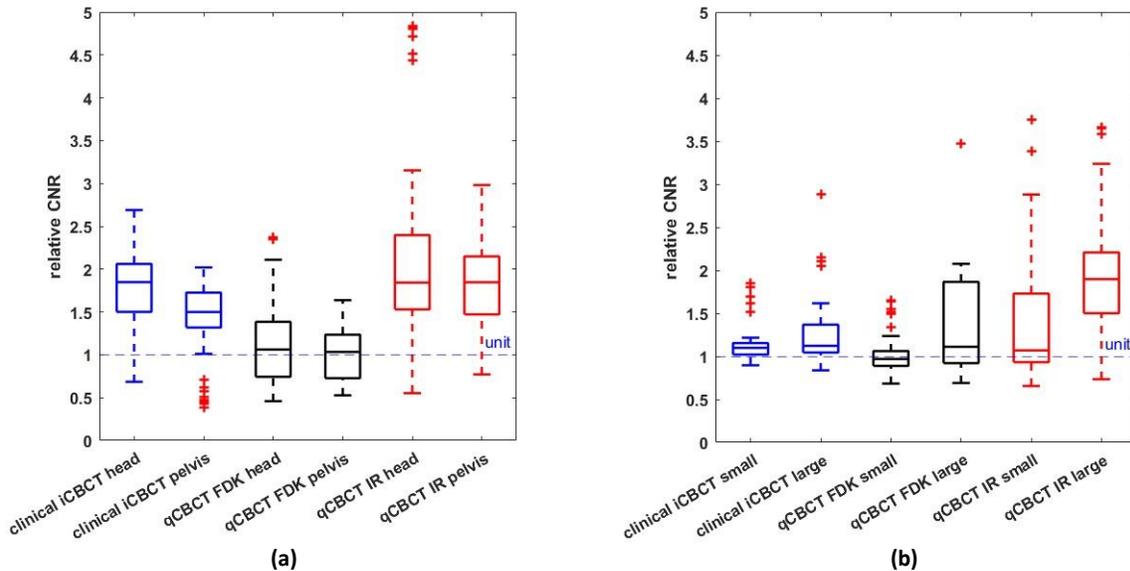

**(a)**                                        **(b)**

*Fig.11. Boxplots of rCNR as a function of CBCT imaging methods in (a) head and pelvis-sized electron density phantom and (b) small and large Catphan phantoms. Relative CNR is the change in CNR with respect to Clinical FDK reconstructions.*

Mean rCNR in Clinical iCBCT images was in range of 1.1 - 1.8 head in electron density and Catphan phantoms (Fig. 11). Mean rCNR in qCBCT FDK images was in the range of 1-1.4 respectively, indicating that qCBCT FDK and Clinical FDK images provide comparable CNRs. rCNR in qCBCT IR images were 1.4 - 2.4, indicating that qCBCT can offer substantial CNR improvement using OS-ASD-POCS reconstruction algorithm. Fig. 12 showcases how using OS-ASD-POCS for qCBCT reduces noise for different phantoms compared with qCBCT FDK. Overall, a noise reduction factor of 2.35 was achieved.

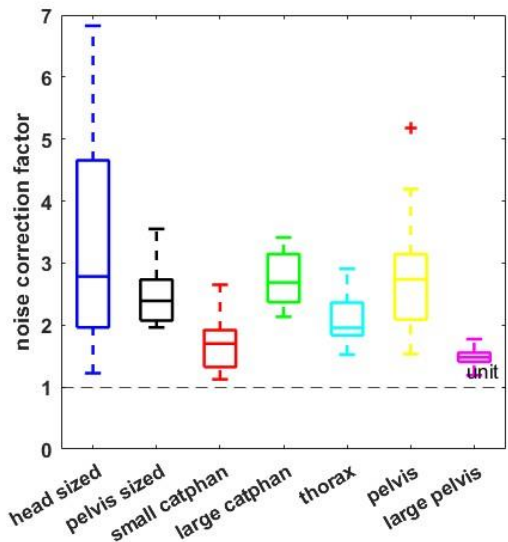

*Fig.12. Boxplot of noise reduction factor for qCBCT IR with respect to qCBCT FDK.*

### 3.2.3. HU nonuniformity

Similar to the trends in HU loss, HU nonuniformity was comparable in head sized phantoms among all imaging methods. HU nonuniformity was larger in pelvis-sized phantoms with high density inserts where effects of SPR, beam hardening, and image lag on HU nonuniformity are larger.

When compared to Clinical CBCT images, HU nonuniformity was substantially less in qCBCT images, and particularly in pelvis-sized phantoms (Table 1 and Fig. 13); mean nonuniformity across all phantoms was 32, 23, 10, 10, and 6 HU for Clinical FDK, iCBCT, qCBCT FDK, qCBCT IR, and MDCT images, respectively. However, large ROI-to-ROI HU deviations were observed in a small subset of ROIs due to image artifacts. Mean HU nonuniformity statistics do not reflect such HU deviations in CBCT images. For example, maximum ROI-to-ROI HU deviation was 193 HU in Clinical FDK images of the pelvis electron density phantom. Such ROI-to-ROI deviations were only 15 HU in qCBCT FDK images.

| | Head electron density | Pelvis electron density | Small Catphan | Large Catphan | Thorax | Pelvis | Large pelvis |
|---|---|---|---|---|---|---|---|
| **Clinical FDK** | 20±14 | 75±57 | 6±4 | 29±21 | 38±29 | 18±11 | 67±31 |
| **Clinical iCBCT** | 13±11 | 32±20 | 4±2 | 19±12 | 30±19 | 21±13 | 49±32 |
| **qCBCT FDK** | 4±4 | 8±3 | 6±2 | 4±4 | 6±4 | 14±10 | 26±18 |
| **qCBCT IR** | 4±4 | 5±3 | 2±1 | 4±3 | 6±4 | 14±10 | 31±20 |
| **MDCT** | 8±4 | 4±3 | 0.4±0.5 | 2±1 | 3±2 | 9±7 | 17±12 |

*Table 1. HU nonuniformity for all phantoms and imaging methods. Mean ± std.*

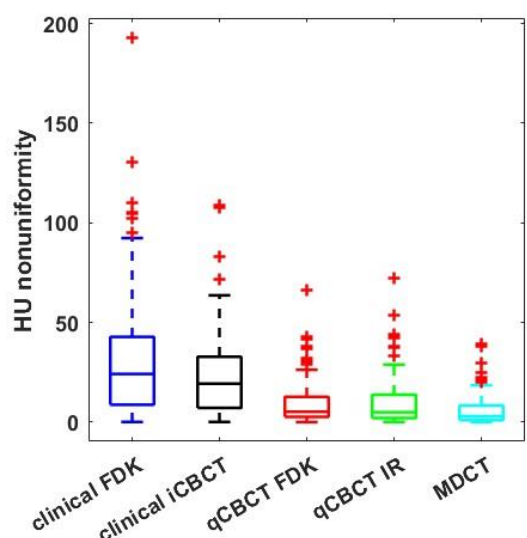

*Fig.13. Boxplots for HU nonuniformity as a function of CBCT imaging methods. Centerline, box, and whiskers represent median, 25-75 quartile, and most extreme data points not considered outliers, respectively. The outliers are presented by crosses.*

### 3.2.4. Analysis of spatial resolution characteristics

Overall, spatial resolution properties were similar among Clinical CBCT and qCBCT images, and spatial resolution was slightly less in the MDCT image (Fig.14a). Values of MTF at 10% level for Clinical FDK, iCBCT, qCBCT FDK, and qCBCT IR were 0.39, 0.41, 0.43, and 0.42 mm$^{-1}$, respectively. 10% MTF level in MDCT images was 0.37 mm$^{-1}$.

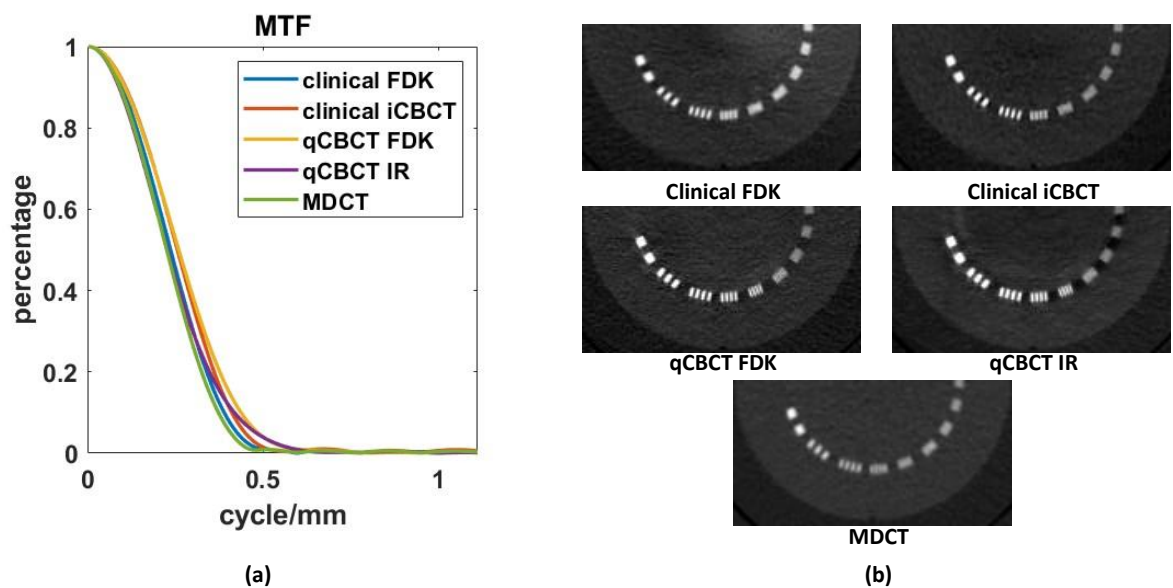

Fig.14. Comparison of spatial resolution for different imaging methods. (a) MTF and (b) bar pattern phantom images.

In bar pattern phantom images, 4 and 5 line pairs per cm were differentiated with Clinical CBCT and qCBCT images, respectively. Whereas only 3 line pairs per cm could be distinguished in MDCT images.

### 3.2.5. HU correlation histograms

Trends in CBCT-MDCT HU correlation histograms agreed with HU loss and nonuniformity evaluations. In soft tissue-mimicking regions of head sized phantoms, HU values for CBCT and MDCT were highly correlated due to highly accurate raw data and lower noise (Fig. 15).

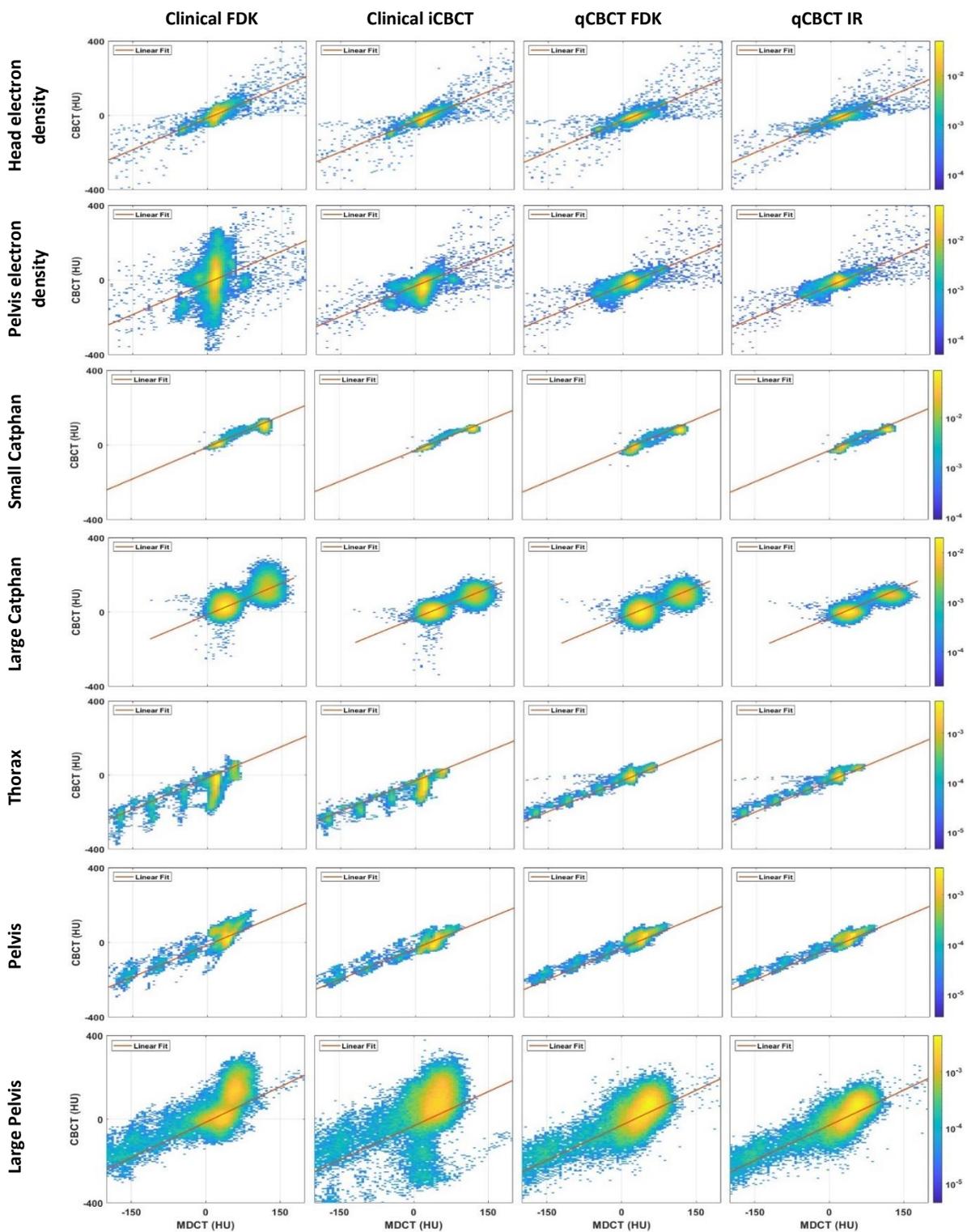

*Fig.15. HU correlation histograms for all phantoms and respective CBCT methods. Red line is the linear fit to CBCT-MDCT HU values in head sized electron density phantom images, representing the ideal CBCT-MDCT HU correlation.*

For larger phantoms, correlation between CBCT and MDCT images degraded as indicated by the dispersion of histogram values from the ideal correlation indicated by the red line. This is not only due to contamination of projection signal with scatter, beam hardening and image lag, but also due to increased noise. h-RMSE (Table 2) values were in the range 16 – 57 HU for Clinical CBCT images, whereas h-RMSE was in the range of 7-31 HU for qCBCT images. With iterative reconstruction, correlation between qCBCT and MDCT was improved modestly due to reduced image noise. Merging all HU correlation histograms in Fig. 15 into one composite histogram better visualizes the dispersion of CBCT HU values (Fig.16).

| | Head electron density | Pelvis electron density | Small Catphan | Large Catphan | Thorax | Pelvis | Large pelvis |
|---|---|---|---|---|---|---|---|
| **Clinical FDK** | 10 | 31 | 9 | 17 | 32 | 16 | 40 |
| **Clinical iCBCT** | 8 | 19 | 6 | 13 | 27 | 13 | 57 |
| **qCBCT FDK** | 7 | 10 | 13 | 17 | 8 | 13 | 31 |
| **qCBCT IR** | 6 | 9 | 12 | 13 | 7 | 12 | 25 |

*Table 2. h-RMSE values in soft tissue mimicking regions of CBCT images when compared to gold standard MDCT images.*

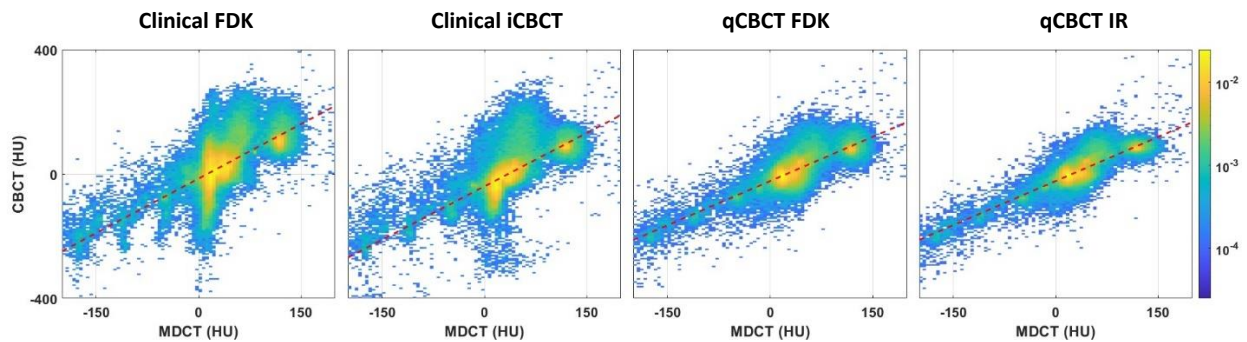

*Fig 16. Composite HU correlation histograms over all phantoms for each CBCT imaging method. Histograms were generated from soft tissue mimicking regions of each phantom.*

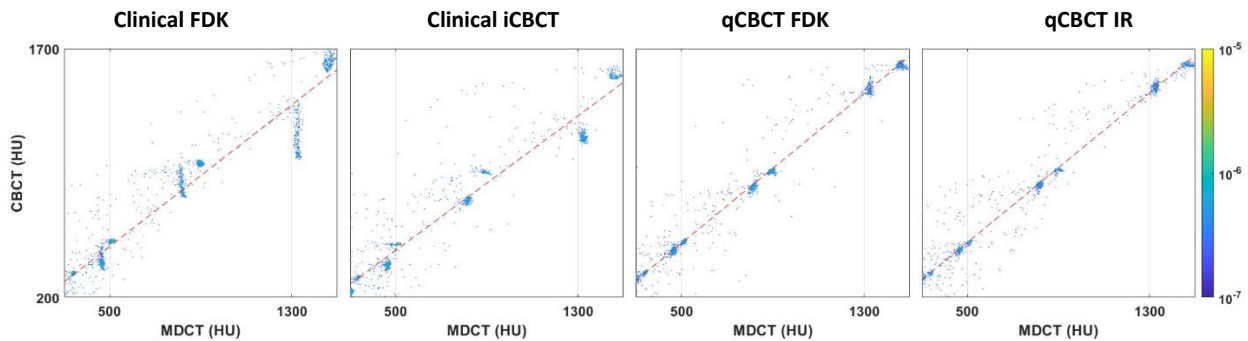

*Fig 17. Composite HU correlation histograms over all phantoms for each CBCT imaging method. Histogram were generated from bone mimicking regions of each phantom.*

HU correlation degraded substantially for bone-like regions in Clinical FDK images (Fig. 17), indicating the challenges in improving raw data fidelity in high SPR conditions. qCBCT images preserved the expected linear behavior of HU values when compared to Clinical CBCT images. h-RMSE values for Clinical FDK, iCBCT, qCBCT FDK, and qCBCT IR were 23, 19, 15, and 12 HU for soft tissue-mimicking regions and 56, 52, 21 and 21 HU for bone-mimicking regions respectively. While noise reduction in qCBCT IR helped, improvement of raw data fidelity was the key factor in achieving higher qCBCT-MDCT HU correlation.

## 4. Discussion

While the performance of 2D grids in CBCT imaging has been investigated previously [6,8,32], HU accuracy and CNR improvement was limited due to lack of key raw data correction and image denoising methods.  In this work, we presented a raw data correction and iterative reconstruction pipeline with 2D antiscatter grids, referred to as qCBCT, which further improves quality of CBCT images.

In qCBCT images, a substantial reduction in image artifacts and improvement in HU accuracy were observed due to a combination of scatter, image lag, and beam hardening correction. Even though scatter rejection and correction are the primary factors in achieving high image quality in qCBCT, they are not sufficient to achieve high HU accuracy comparable to MDCT. For example, localized HU nonuniformities in the periphery of the pelvis electron density phantom were 15 to 46 HU due to image lag and beam hardening.  After beam hardening and image lag correction, HU nonuniformity was about 8 HU on the average in qCBCT images, which points to the essence of complete raw data correction chain in achieving quantitatively accurate CBCT images.

Regarding HU accuracy, qCBCT provided significant improvements over both Clinical CBCT imaging methods investigated. Since both qCBCT and Clinical CBCT utilize beam hardening and image lag correction methods, improved image quality in qCBCT was attributed to 2D antiscatter grids. Even though 2D grid does not eliminate all scatter, the remaining smaller amount of scatter can be more easily corrected with a scatter correction method. Conventional radiographic antiscatter grid in Clinical CBCT transmits a larger fraction of scatter fluence to the detector and makes residual scatter correction a more challenging task. Our results also reaffirm that iCBCT has significantly better scatter correction performance than the scatter kernel superposition-based method used in Clinical FDK images.

HU accuracy of qCBCT and MDCT images were comparable over aggregate evaluations across all ROIs and phantoms. Yet, MDCT provided slightly less artifacts, particularly in larger phantoms. Such HU accuracy differences, albeit small, might be attributed to suboptimal scatter suppression, beam hardening, and lag correction. Another contributing factor might be the limitations of FPD technology. For example, primary x-ray fluence can be attenuated more than 5 orders of magnitude in large, bone-containing phantoms. At such low signal levels, limited dynamic range and relatively high electronic noise in FPDs [34] may challenge raw data correction and accurate measurement of primary signals. These issues are further amplified by relatively low quantum efficiency of FPDs and associated quantum noise in primary signal.

CNR values were comparable between qCBCT and clinical CBCT images that were reconstructed using FDK. Given that the 2D antiscatter grid in qCBCT has higher primary transmission than the radiographic grid used in clinical CBCT (85% versus 70%) [8,32], and 2D grid provides better scatter rejection performance, one would expect to achieve higher CNR in qCBCT images. Several factors might have contributed to lower-than-expected CNR performance in qCBCT FDK images. First, measured CNR values are sensitive to spatial resolution properties. qCBCT FDK images had slightly higher spatial resolution than Clinical CBCT images (Fig.14a). If spatial resolution properties of qCBCT and Clinical CBCT images are matched better, CNR in qCBCT images can be improved further. Second, qCBCT images did not employ any high spatial frequency artifact reduction algorithms, such as ring artifact correction methods, as in Clinical CBCT. Such high frequency artifacts can increase noise and reduce CNR. Implementation of such artifact reduction methods in qCBCT will be an area of future investigations.

Scatter rejection by 2D grid and residual scatter correction with the GSS method improve contrast but also amplify noise in CBCT images [6]. Hence, noise reduction strategies have a larger potential to improve low contrast visualization and CNR in qCBCT. OS-ASD-POCS iterative reconstruction method provided significant reduction in noise and on the average 72% increase in CNR. Despite noise reduction, qCBCT images of large phantoms appear noisier than MDCT and on par with Clinical CBCT, which imply the need for improved noise reduction in qCBCT.

Three different CT number accuracy metrics were utilized, as each method has its own advantage and drawback in capturing the full extent of CT number degradation. For example, HU loss metric is powerful in characterizing HU bias due to change in phantom dimensions, but evaluations are limited to the volume of the small phantom used in evaluations. HU nonuniformity metric is limited to evaluation of HU deviations in uniform regions of phantoms. CBCT-MDCT HU correlation histogram can potentially characterize HU inaccuracy in a more comprehensive fashion across all image voxels. However, accuracy of HU correlation histograms was limited by the inherent CT number accuracy of MDCT images. Therefore, utilization of multiple metrics yields a more complete picture of CT number accuracy in CBCT images.

This work emphasizes the importance of phantom size and composition in CBCT image quality evaluations. Image quality differences among all imaging methods evaluated were generally small in 20 cm diameter phantoms, because the effects of scatter, beam hardening and image lag are less evident in such phantoms in linac-mounted CBCT geometry. When human torso sized phantoms with bony structures were used, HU loss exceeded 100 HU, implying a drastic reduction in quantitative accuracy. Hence, the use of standard head-sized phantoms and absence of bone-like embedded objects in CBCT image quality evaluations may not fully reveal the extent of image quality issues in CBCT images [16,35].

## 5. Conclusion

Utilization of 2D antiscatter grid in conjunction with raw data correction and iterative reconstruction pipeline can provide significant improvements in quality of CBCT images used in radiation therapy. As demonstrated in a variety of phantoms, HU accuracy in qCBCT images approaches the HU accuracy of gold standard MDCT images. CNR in iteratively reconstructed qCBCT images is also improved, promising better soft tissue visualization.

From a qualitative perspective, improved image quality in qCBCT may increase clinician's confidence during target localization in image guided radiation therapy. From quantitative imaging perspective, qCBCT images can be potentially used for CBCT-based dose calculations during treatment delivery to either confirm delivered dose or support decision for plan adaptation. Furthermore, anatomical changes and treatment response over the course of treatment can be assessed better in qCBCT images.

**Acknowledgements**

This work was funded in part by grants from NIH/NCI R21CA198462 and R01CA245270.

**Conflict of Interest Statement**

The authors have no relevant conflicts of interest to disclose.